\documentclass{elsart}
\usepackage{epsfig}
\usepackage{amssymb}
\usepackage{rotating}

\begin{document}

\begin{frontmatter}

\title{Reply to comments by Benoit \textit{et al} on the ZEPLIN I dark matter limit}

\author[RAL]{N. J. T. Smith\corauthref{cor1}},
\corauth[cor1]{Corresponding author; address: Particle Physics Department, CCLRC Rutherford
Appleton Laboratory, Chilton, Didcot, Oxon, OX11 0QX}
\ead{n.j.t.smith@rl.ac.uk}
\author[EDI]{A. S. Murphy},
\author[ICL]{T. J. Sumner}

\address[RAL] {Particle Physics Department, CCLRC Rutherford Appleton Laboratory, UK}
\address[EDI] {School of Physics, University of Edinburgh, UK}
\address[ICL] {Blackett Laboratory, Imperial College London, UK}

\begin{abstract}
Recent objections ({\it Phys.Lett. B} {\bf 637}, 156.) to the published Zeplin I limit ({\it Astropart. Phys} {\bf 23}, 444)  are shown to arise from misunderstandings of the calibration data and procedure, and a misreading of the data in one of the referenced papers.  
\end{abstract}


\end{frontmatter}

% Main text %

\section{Introduction}

Following our publication of a dark matter limit from the UK ZEPLIN I detector \cite{zeplin1} ,  A Benoit  \textit{et al.} \cite{benoit} have claimed the time constant calibration to be unsound and inconsistent with other published data \cite{akimov} and hence that the limit is invalid.  We consider their points in turn:

\subsection{Event separation}
Benoit  \textit{et al.} state their opinion that discrimination between background and signal can only be achieved using completely separated distributions.  This approach is fundamentally incorrect; no two distributions ever have  zero overlap and thus one can only ever make a statistical statement as to the degree of separation of the two distributions as a function of the recoil energy. Different techniques may need to be applied to extract the underlying nuclear recoil population depending on this statistical statement, the technique used in the ZEPLIN I analysis is detailed in our paper  \cite{zeplin1}.

\subsection{Quenching factor}
In converting our visible energy range into recoil energy Benoit  \textit{et al.}, choose to use the results of Aprile \textit{et al.} \cite{aprile} not available at the time of the original paper.  However it should be noted that there are now additional measurements in the literature \cite{chepel} at  even lower energy that report higher values. Although this discrepancy has not been  resolved, the effect of varying the quenching factor on the dark matter limit is fully detailed in the systematics review within our paper.

\subsection{The time constant data of Akimov et al.\cite{akimov}}
In a footnote in \S6 of our paper, we draw attention to the fact that there is no inconsistency between \cite{akimov} and our data because \cite{akimov} does not provide nuclear recoil data below 10 keV.   Benoit \textit{et al.} reproduce Figure 4 of \cite{akimov} but unfortunately do not reproduce its original caption, which states clearly that the points shown at 7 keV and 15 keV do not refer to those energies, but apply to the whole energy range from 6 to 30 keV.   The Akimov  \textit{et al.}. caption also explains that this wide horizontal error bar was omitted from their plot Òfor clarityÓ.   For this reason those results are, as we stated in \cite{zeplin1}, are of very little relevance to the ZEPLIN I analysis.

\subsection{The ambient run}
Our data are reproduced by Beniot  \textit{et al.} in their figure 3.  Our paper clearly states that this was shown as corroborative data, but not used in the analysis, since the neutron-like pulses, though identical to neutron pulses in shape, cannot be proved to be unambiguously from neutrons.  

\subsection{Event rates for AmBe and ambient runs}
These runs are not comparable as ``source present'' and ``source removed''.   The AmBe runs had been intended as a preliminary calibration only, and had a running time two orders of magnitude shorter than that of the ambient run.  Thus for this shorter running time the number of neutron events expected from the ambient population is entirely negligible; the neutrons are thus from the AmBe source. 

\subsection{AmBe run significance}
Benoit  \textit{et al.}claim that visual examination of the tail of our AmBe exposure distribution shows no neutron events.  However, this must be compared with the shape of the
$\gamma$-ray only time constant distribution also shown in our figure.   This is found from high-statistics $^{60}$Co and $^{137}$Cs  Compton scattering calibration runs to have a very sharp fall-off of the left hand branch.  As discussed in our paper, this has been determined with high accuracy and fits a $\gamma$-ray  distribution at all energies.  We show one example of this in Figure 2 of our paper.   A statistical analysis of the AmBe source population, together with the high accuracy Compton scattering runs, then shows a clearly significant additional population on the tail when the AmBe source is in place. From these data the time constant ratio can be determined with the error bar which we quote.   Plotting the data in logarithmic or linear scale makes no difference to the statistical significance of the short time-constant population.

\subsection{Neutron fraction}
Benoit  \textit{et al.} also suggest a discrepancy in the neutron fractions in our 20-30 keV and 3-8 keV fits.   However, although perhaps not obvious from the original paper, the two cases shown are from different runs with different CsI tagging efficiencies, as it was necessary to use different DAQ gain and tagging parameters to obtain data sets above and below 10 keV. The absolute event numbers do not, of course, affect the time constant ratio, with the statistical influence of the small number of nuclear recoils at low energy accounted for as described in our paper.   

\subsection{Efficiency of noise cut}
The energy-dependent efficiency of the photomultiplier detector noise cut was first estimated, from simulations, to be in the region 70-90\%.  As stated in our paper, we were able to confirm this by comparing the fitted AmBe neutron source response in various energy intervals below 10keV with and without applying the noise cut.  It is these experimental efficiencies which are used in our Table 2.    It should also be noted that, for the second population observed in both the AmBe neutron source runs and ambient runs, the individual pulses were examined and found to be of the shape characteristic of particle events, and not the shape observed for electronic or photomultiplier tube noise events.

\subsection{Variations in absolute time constant}
The ambient and AmBe source runs were not in fact contemporary Òsource inÓ and Òsource outÓ measurements as assumed by Benoit  \textit{et al.},  but carried out at different times during several months of laboratory testing, and with different Xe fills.  Since the absolute time constant has some dependence on Xe purity, some variation in time constant is possible from run to run, in addition to the monotonic decrease of absolute time constant with energy observed in all runs (as noted by Benoit  \textit{et al.} for our high and low energy AmBe source runs). As stated in our paper, the intention was to perform a full underground neutron/gamma calibration with a similar detector configuration as for the underground data. However, this was curtailed due to a technical failure within the thermal control mechanism.

\subsection{Gamma rejection factor}
Benoit  \textit{et al.} express scepticism at the large background rejection factor achievable with ZEPLIN I.   This good background rejection results from the fact that any signal would lie at the tail of the $\gamma$-ray  $\tau$ distribution, where the latter is shown by Compton scattering calibration data to be falling very sharply at all energies, and with entirely positive curvature in the quantity d(log N)/d($\tau$). The majority of the $\gamma$-ray (and $\beta$ particle) background lies outside the signal region and part of the signal region is indeed extremely well separated from the $\gamma$-ray population.  Thus analysis is performmed using Poisson statistics applied to a region with very few $\gamma$-ray background events (typically 0 - 2 in a 2keV energy range).  This is illustrated in detail by Figures  4 and 5 of our paper, which show how, in a given energy range, a signal limit of ~10 events could indeed be seen as a significant slope change at the tail of a population of 10000 $\gamma$-ray events.   The significance and sensitivity then continue to improve with running time.   Thus, this point of concern in\cite{benoit} is already discussed in detail in \S8 of our paper.

\section{Final Comments}

It remains the case that it was not the original intention to use the preliminary laboratory neutron run as a final calibration.  As stated in our paper, the planned high-statistics neutron calibration of ZEPLIN I was precluded by curtailment of the underground runs due to a technical failure which would have necessitated removal and significant refurbishment of the detector.   Resources could not be allocated to this because of the higher priority need to complete the new more powerful hybrid detectors, ZEPLIN II and ZEPLIN III. Accordingly we have set a dark matter limit using the preliminary calibration data and a conservative time constant ratio from those data, with a correspondingly large error bar stated in our paper.  This error bar was folded into the subsequent analysis to give a limit which was poorer than would have been the case with a more accurate neutron calibration.  Our paper discusses in detail the impact of various systematic errors (e.g. quenching factor) on this final limit.  More recent measurements reported by the Columbia and Coimbra groups \cite{aprile,chepel} show that the quenching factor does sustain values in the region 0.2 $\pm$0.05 down to $\sim$10 keV, so that this assumption now appears to be valid. 

We have shown that the objections of Benoit  \textit{et al.} to the ZEPLIN I dark matter limit arise from misunderstandings of the calibration data and procedure, and a misreading of the data in one of the referenced papers. Accordingly, we  believe the statistical extraction of the limit to be sound, and not over estimated by a factor of a 1000 as stated by Benoit  \textit{et al.}, with a  detailed systematics analysis given in the appendix of our paper.

\end{document}